\documentclass[12pt]{aastex63}
\usepackage{graphicx}
\usepackage{geometry}
\shorttitle{Observations of SDSS~J1004+4112}
\shortauthors{Jackson}
\begin{document}
\title{The faintest radio source yet: EVLA observations of the
gravitational lens SDSS~J1004+4112}
\author{N. Jackson}
\affil{Jodrell Bank Centre for Astrophysics, School of Physics \& Astronomy, 
University of Manchester, Alan Turing Building, Oxford Road, 
Manchester M13 9PL, UK}

\begin{abstract}
We present new radio observations of the large-separation gravitationally-lensed
quasar SDSS~J1004+4112, taken in a total of 6 hours of observations with the Extended Very
Large Array (EVLA). The maps reach a thermal noise level of approximately 7$\mu$Jy. 
We detect four of the five lensed images at the 30-65$\mu$Jy level, representing
a source of intrinsic flux density, after allowing for lensing magnification, of 
about 2$\mu$Jy, intrinsically probably the faintest radio source yet detected. This 
reinforces the utility of gravitational lensing in potentially
allowing us to study nanoJy-level sources before the advent of the SKA. In an 
optical observation taken three months after the radio observation, image C is 
the brightest image, whereas the radio map shows flux density ratios consistent 
with previous optical observations. Future observations separated by a time 
delay will give the intrinsic flux ratios of the images in this source.
\end{abstract}


\section{Introduction}

Quasars which are gravitationally lensed by foreground galaxies are important for 
many different topics in astrophysics and cosmology. These range from 
determination of the Hubble constant (Refsdal 1964; see e.g. Kochanek \& Schechter 
2004; Jackson 2007 for reviews) to the study of the individual
lens systems. These studies can be divided into studies of the lensing galaxies, 
in particular the determination of their dark-matter content and distribution; 
and studies of the quasars, in which we can probe intrinsically faint objects 
because of the lensing magnification.

Studies of the lensing galaxies use the positions and flux densities of the 
lensed quasar images, which probe the gravitational potential at points where 
the corresponding light ray cuts the lens plane. In favourable cases, constraints 
on the lens potential can be obtained, notably in the case of CLASS~B1933+503 
where three components of the background object are lensed (Sykes et al. 1998; 
Nair 1998; Cohn et al. 2001).Often, however, it is found that smooth models fail
to reproduce the observed image fluxes. This was first noted by Mao \& Schneider (1998) in
the case of CLASS~B1422+231, and subsequently the effect was studied in samples
of quadruple quasar lens systems (Dalal \& Kochanek 2002, Kochanek \& Dalal 2004, 
Metcalf 2002, Chiba 2002). The lack of a good fit is often ascribed to the presence
of dark substructure, on scales ranging down to $10^6M_{\odot}$, which is predicted
by CDM simulations (e.g. Diemand et al. 2008). Fluxes are more sensitive to small-scale 
irregularities in the mass field than image positions, because they are dependent on 
the second rather than the first derivative of the potential, although sometimes
accurate position information cannot be well fit (e.g. CLASS~B0128+437; Phillips
et al. 2000, Biggs et al. 2004). The predicted substructure is apparently not seen 
in our own Galaxy, a phenomenon known as the ``missing-satellites problem'' (Moore 
et al. 1999, Klypin et al. 1999), and it may be that star formation in Galactic 
satellites is suppressed (Bullock et al. 2000). In lensing galaxies, the ``substructure''
detected through lensing actually exceeds that predicted by CDM, because in the
central regions where lensing constraints are available, the substructure fraction
is expected to be $<1$\% (Mao et al. 2004; Xu et al. 2009).

The investigation of flux anomalies in lens systems is currently plagued by 
small-sample statistics, because the most suitable lens systems for
study are those relatively small numbers of objects where radio, or other low-frequency, 
measurements are currently obtainable. The low-frequency emission comes from regions of 
the source which are relatively extended, and consequently are not subject to 
microlensing by the stars in the intervening galaxy, which affects the optical 
fluxes (e.g. Schechter \& Wambsganss 2002), or by optical extinction effects. 
The only effects present, if microlensing is excluded, are the effects of the 
mass distribution of the lens, together with the combined effect of variability 
in the source together with relative time delays in the images. There may also 
be mild effects of scattering (Koopmans et al. 2003). Unfortunately, however, only 
a dozen radio-loud, quadruply imaged quasars are known, mostly from the CLASS survey 
(Myers et al. 2003, Browne et al. 2003) but also from deep radio images of
less radio-loud sources (e.g. Kratzer 
et al. 2011). Many authors have attempted to use mid-infrared fluxes instead
(Chiba et al. 2005, Fadely \& Keeton 2010) as an alternative waveband to study radio-quiet
quadruple lens systems, and several further flux-anomalous systems have been detected
in this way. In the future, however, the advent of very sensitive radio interferometers
such as the EVLA and e-MERLIN, which have microJansky sensitivity levels coupled with
sub-arcsecond resolution, will allow study of hitherto ``radio-quiet'' radio sources. 
It has been shown by stacking of images from the FIRST survey (White et al. 2007) that
typical radio-quiet quasars of optical $I$ magnitudes of 18-20 should have radio 
flux densities of a few tens of $\mu$Jy up to $\sim$150$\mu$Jy, very 
suitable for studies with the new radio arrays. In principle, nearly a hundred new
lens systems with known radio flux densities could be found by radio follow-up of
known radio-quiet lens systems.

Studies of the sources are also potentially rewarding, because the lensing
magnification allows us to study quasars at flux density levels which we would
otherwise not be able to reach. For example, it is not yet known whether the
radio emission mechanism in radio-quiet quasars is similar to that in radio-loud
quasars, with a compact core in which jet-like emission is collimated, or whether
some other mechanism such as optically-thin free-free emission is at work
(Blundell \& Kuncic 2007). Detailed studies of the radio emission, or comparison
of variability properties in radio and optical, may help here. However, the
radio-faintness of many radio-quiet quasars is a challenge even for modern
radio arrays such as the EVLA, and routine observations of these objects
may be greatly assisted by choosing a lensed sample containing quasars which
are magnified, typically by factors of 5--10.

\section{The gravitational lens system SDSS~J1004+4112}

As a beginning to such a programme, we present new, deep EVLA observations of the lensed
quasar SDSS~J1004+4112. This system consists of a $z_s=1.73$ quasar being lensed by
a galaxy cluster at $z=0.68$ into five images, with a maximum separation of 14\farcs6,
and was discovered in the Sloan Quasar Lens Survey (Inada et al. 2003; Oguri et al. 2004; 
Inada et al. 2005) using the catalogue of quasars from the Sloan Digital Sky Survey 
(Schneider et al. 2007). It has been modelled by numerous authors (Williams \& Saha 2004, 
Kawano \& Oguri 2006, Saha et al. 2007, Fohlmeister et al. 2007, 
Liesenborgs et al. 2009, Oguri 2010) using constraints including multiple time delays
(Fohlmeister et al. 2007, 2008), spectroscopy of galaxies in the cluster (Sharon et al.
2005), and {\em Chandra} X-ray observations (Ota et al. 2006). The magnifications of
the images are likely to be considerable, with the exception of the central image, E: 
Oguri (2010) estimates the image magnifications to be 29.7, 19.6, 11.6, 5.8 and 0.16 
for A, B, C, D and E respectively. Optical microlensing
is known to exist in this system (Fohlmeister et al. 2008) and has been used to determine an 
approximate size for the accretion disk in the source quasar.

\section{Observations and results}

Observations were obtained on 4 epochs: 2010 October 15, 2010 November 
15, 18, 20, using the Extended Very Large Array (EVLA) in C-configuration. 
Each observation
consisted of 9$\times$370-s scans within a total observing time of 90
minutes, interspersed with observations of a phase calibrator 
(J0948+4039). This resulted in a total time on source of just under
4 hours. 3C286 was used as an absolute flux calibrator (Baars et al. 1977).
All observations were carried out in two contiguous, 128-MHz IF bands, each
divided into 64 channels, and centred at 4896 and 5024~MHz. 

The data were processed in the NRAO {\sc aips} package. Significant phase
slopes across the bandpass were present, which were corrected by fringe
fitting to the 3C286 observations. The resulting delay and rate solutions,
consisting of delays of typically a few ns, were applied to the phase
calibrator J0948+4039 to check their validity. A few channels at the edge
of each IF were deleted, and a bandpass solution was made, again using
3C286. Phase calibration solutions were then derived using J0948+4039. 
In some epochs the atmospheric phase varied by up to a radian during the
observation, but this could be followed well by the phase calibration
observations. Images were made using the {\sc aips imagr} routine using
Briggs {\sc robust}=0 weighting; natural weighting was also attempted but
this produced no noticeable improvement in the signal-to-noise while
degrading the beam considerably. The final images have a resolution
of 3\farcs 95 $\times$   3\farcs 69 in position angle 58$^{\circ}$. and
the off-source noise level is approximately 7$\mu$Jy/beam.

The cleaned map is reproduced in Fig. 1, superimposed upon an archival
HST image (GO-10509, PI Kochanek). The four bright images of
the quasar (A-D) are all clearly detected, although the A and B images are 
only marginally resolved from each other. All the radio components have 
flux densities between 30 and 65 $\mu$Jy (Table 1). No radio emission is 
seen from the main lensing galaxy G, which is also known to contain a faint fifth 
optical image E. A much brighter radio source, R, is seen about 25\arcsec
north of G. This has a flux density of 2.7~mJy, but is only marginally 
visible, at about 0.5~mJy, in the FIRST 1.4-GHz image (Becker, 
White \& Helfand 1995). It is therefore either highly variable, or else an
inverted-spectrum source.

\begin{figure*}
\plotone{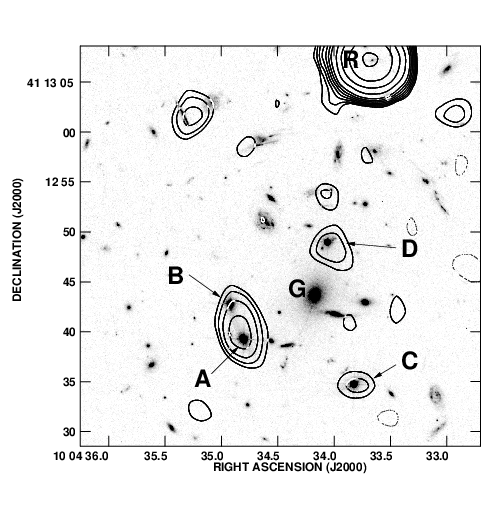}
\caption{EVLA 4.959-GHz
radio contours superimposed on the archival HST image of the 
SDSS~J1004+4112 field. Contours are given with a base level of 
18.5$\mu$Jy/beam (approximately 2.5$\sigma$) with multiples $-$1 (dotted),
1,1.41,2,2.82,4,5.6,8,16,32,...,1024.}
\end{figure*}

Optical imaging observations of J1004+4112 were also made using the 4.2-m 
William Herschel Telescope on La Palma, on the night of 2010 February 25.
The ACAM camera was used and images were obtained in two colours, corresponding
to the Sloan $g$ and $r$ filters with 2$\times$300s exposure in each filter. 
The $r$ image is presented in Fig. 2 and
the image flux densities, together with the EVLA radio flux densities, are
shown in Table 1. 

\begin{table}
\caption{Flux densities of the four bright components of SDSS~J1004. EVLA 5-GHz flux densities
are given in $\mu$Jy, together with the magnification $\mu$ predicted for each image
in the model of Oguri (2010). The WHT/ACAM optical fluxes in Sloan $g$ and $r$ magnitudes
are also given in the last two columns.}
\begin{tabular}{ccccc} \hline \hline
Cpt.& EVLA & $\mu$ & Sloan $m_g$ & Sloan $m_r$ \\ 
    & F$_{\rm 5GHz}$/$\mu$Jy &&&\\ \hline
A & 64$\pm$8 & 29.7 & 20.83$\pm$0.03 & 20.27$\pm$0.03 \\
B & 39$\pm$8 & 19.6 & 21.26$\pm$0.03 & 20.91$\pm$0.03 \\
C & 30$\pm$8 & 11.6 & 20.13$\pm$0.03 & 20.03$\pm$0.03 \\
D & 33$\pm$8 &  5.8 & 20.99$\pm$0.03 & 20.79$\pm$0.03 \\ \hline
\end{tabular}
\end{table}

\begin{figure}
\epsscale{0.9}
\plotone{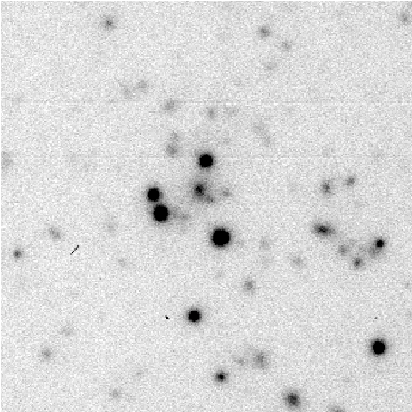}
\caption{WHT $r$-band image of J1004+4112, taken on the night of
2011 February 25. Note the relatively much brighter
image C compared to the radio image in Fig. 1.}
\end{figure}

\section{Discussion and conclusions}

The current observations are the first of a programme which
may tell us much both about the lenses and the lensed quasar. The lens can
be probed by knowing the intrinsic long-wavelength flux ratios, together
with other constraints from previous observations. The nature of
the quasar's radio source can be probed by its flux density at 
different wavelengths, and also by its variability properties in the 
optical and radio wavebands.

\subsection{Flux ratios and the lensing galaxy}

In principle, radio observations tell us the intrinsic flux ratios without
interference from microlensing effects or extinction. If the radio source
is not variable, then the radio flux ratios are consistent with models 
(see Table 1) except for component D, which is relatively bright. 
The picture can be complicated by the combined effects of variability 
and time delay, and in this case also by possible differences in variability
characteristics in the two wavebands. 

Fohlmeister et al. (2008) present optical monitoring of SDSS~J1004+4112 for
nearly four years from the end of 2003 until the middle of 2007. Flux
variations in the five images are expected to proceed in the order C-B-A-D-E,
where the C-A and B-A time delays have been measured by Fohlmeister et al. (2008) 
as 821.6$\pm$2.1 days and 40.6$\pm$1.8 days, respectively; they also obtained a lower
limit on the A-D delay of 1250 days. Model predictions for this delay include
1218 days by Oguri et al. (2010) and $\sim$2000 days by Fohlmeister et al.
(2008). Fohlmeister et al. obtain delay-corrected 
flux ratios of B to A that vary from 0.283 to 0.460 mag due to microlensing, 
and in C to A of 0.59 mag. Comparison of the WHT optical photometry of 2011 
February with Fohlmeister et al. (2008), whose photometry covers the period 
2004-07, implies that components A and B have continued to decline in brightness 
by about half a magnitude in the last three years. It will be important to
use radio observations to derive intrinsic flux ratios free from the effects
of microlensing, using further observations separated by the C-B and C-A time 
delays in the system.


\subsection{Flux level, variability and the lensed object}

The detection of significant radio flux density, albeit at a low level, from
SDSS~J1004+4112 vindicates the prediction of White et al. (2007) that 
hitherto ``radio-quiet'' quasars should display significant radio flux
when imaged with noise levels of a few $\mu$Jy, which are now within reach
using new radio interferometer arrays such as the EVLA and e-Merlin. It also
suggests that the correlation between $I$-band and cm-wave radio flux density
inferred by White et al. continues down to considerably lower flux density
levels than can be probed by FIRST. Moreover,
the high magnification of this lens system implies that the intrinsic flux
of the radio source is between 2 and 4$\mu$Jy, using the model of Oguri (2010).
This is probably the lowest intrinsic flux density of any source yet detected
in the radio. The current faintest radio sources include the lensed submillimetre
galaxy SMMJ16359+6612, detected in deep WSRT observations by Garrett, Knudsen
\& van der Werf (2005) and a lensed radio source in the cluster
MS0451.6$-$0305 (Berciano Alba et al. 2007), both of which have unlensed
flux densities of about 3$\mu$Jy. Similar, or slightly brighter, detections have
been reported in other lensed submillimetre galaxies 
(e.g. Ivison et al. 2010). Other faint radio lenses may emerge from candidates
in the COSMOS field (Faure et al. 2008, Jackson 2008) in which deep radio maps
have been made (Schinnerer et al. 2007) but no other 
lensed quasars are yet reported with such low intrinsic radio fluxes.
Such a radio source, if unlensed, would require about 
a week of observing time using the EVLA when it is fully completed with the 
full 2-GHz bandwidth, in order to achieve a significant detection. In this 
object, we can therefore use gravitational lensing for detailed study of an 
intrinsically weak radio source at a level which will only become routine in 
unlensed sources with the advent of the Square Kilometre Array in the next decade.

One immediate objective that can be resolved soon is the variability of
the radio source. The pattern of the radio source variability depends on the
nature of the radio source, and in particular whether radio-quiet quasars
such as SDSS~J1004+4112 produce radio emission with a standard black-hole/jet,
as in stronger sources, or with a different mechanism such as optically-thin
free-free emission (Blundell \& Kuncic 2007) from a disk wind. In the latter 
case we might expect radio variability to be associate with the variability of the
optical emission from and around the disk. Variability properties of such
faint quasars are unknown, although Barvainis et al. (2005) find similar 
variability properties in samples of radio-loud quasars and quasars with 
radio flux densities $\sim$1~mJy.

A remarkable feature of the present observations is the relative
brightness of C compared to the other components in the 2011 February optical
observations, being then 0.24 mag brighter than A, despite the November
2010 radio flux being less than that of A by a factor of approximately 2.
A 1-magnitude brightening in such a short period has not previously been
seen in optical monitoring data. Either C is currently undergoing a high-
magnitude microlensing event, or a high-amplitude episode of intrinsic 
variability is currently taking place.

\section*{Acknowledgements}
The EVLA is operated by the National Radio Astronomy Observatory, 
a facility of the National Science Foundation operated under cooperative 
agreement by Associated Universities, Inc.
The William Herschel Telescope is operated on the island of La Palma by
the Isaac Newton Group of Telescopes at the Spanish Observatorio del Roque 
de los Muchachos of the Instituto de Astrof\'{\i}sica de Canarias. I thank
Ian Browne and an anonymous referee for comments on the manuscript.

Facilities: EVLA, WHT/ACAM


\bigskip

\hskip 63mm  {\sc REFERENCES}
\bigskip

\scriptsize
\parindent 0mm

Baars, J.W.M., Genzel, R., Pauliny-Toth, I.I.K., Witzel, A. 1977,  A\&A, 61, 99.

Barvainis, R., Lehar, J., Birkinshaw, M., Falcke, H., Blundell, K.M., 2005, ApJ, 618, 108

Becker, R.H., White, R.L., Helfand, D.J. 1995,  ApJ, 450, 559.

Berciano Alba, A., Koopmans, L.V.E., Garrett, M.A., Wucknitz, O., Limousin, M. 2007. A\&A, 509, 54

Biggs, A.D., Browne, I.W.A., Jackson, N., York, T., Norbury, M.A., McKean, J.P, Phillips, P.M., 2004, MNRAS, 350, 949

Blundell, K.M., Kuncic, Z., 2007, ApJ, 668, L103

Browne, I.W.A., Wilkinson, P.N., Jackson, N.J.F., Myers, S.T., Fassnacht, C.D., Koopmans, L.V.E., Marlow, D.R., Norbury, M., Rusin, D., Sykes, C.M.,  2003,  MNRAS, 341, 13.

Bullock, James S., Kravtsov, Andrey V., Weinberg, D.H.2000,  ApJ, 539, 517.

Chiba, M., 2002,  ApJ, 565, 17.

Chiba, M., Takeo, M., Kashikawa, N., Kataza, H., Inoue, K., 2005, ApJ, 627, 53

Cohn, J.D., Kochanek, C.S., McLeod, B.A., Keeton, C.R.2001,  ApJ, 554, 1216.

Dalal, N., Kochanek, C.S.2002,  ApJ, 572, 25.

Diemand, J., Kuhlen, M., Madau, P., Zemp, M., Moore, B., Potter, D., Stadel, J.2008,  Natur, 454, 735.

Fadely, R., Keeton, C.R., 2011, AJ, 141, 101

Faure C., et al., 2008, ApJS, 176, 19

Fohlmeister, J., Kochanek, C.S., Falco, E.E., Wambsganss, J., Morgan, N., Morgan, C.W., Ofek, E.O., Maoz, D., Keeton, C.R., Barentine, J.C.,  2007,  ApJ, 662, 62.

Fohlmeister, J., Kochanek, C.S., Falco, E.E., Morgan, C.W., Wambsganss, J.2008,  ApJ, 676, 761.

Garrett M.A., Knudsen K.K., van der Werf, P.P., 2005, A\&A, 431, L21

Inada, N., et al., 2003, Nature, 426, 810

Inada, N., Oguri, M., Keeton, C.R., Eisenstein, Daniel J., Castander, Francisco J., Chiu, Kuenley, Hall, Patrick B., Hennawi, Joseph F., Johnston, D.E., Pindor, Bartosz,  2005,  PASJ, 57,L7 

Ivison, R.J., et al., 2010, A\&A, 518, L35

Jackson, N., 2007,  LRR, 10, 4.

Jackson, N., 2008, MNRAS, 389, 1311

Kawano, Y., Oguri, M., 2006,  PASJ, 58, 271.

Klypin, A., Kravtsov, A.V., Valenzuela, O., Prada, F., 1999,  ApJ, 522, 82.

Kochanek, C.S., Dalal, N., 2004,  ApJ, 610, 69.

Kochanek, C.S., Schechter, Paul L.2004, Measuring and Modelling the Universe, Carnegie Obs. Cent. Symp., ed. Freedman, W., publ. CUP., p.117.

Koopmans, L.V.E., Biggs, A., Blandford, R.D., Browne, I.W.A., Jackson N., Mao S., Wilkinson, P.N., de Bruyn, A.G., Wambsganss, J., 2003, ApJ, 595, 712

Kratzer, R., et al., 2011, ApJ, 728, L18

Liesenborgs, J., de Rijcke, S., Dejonghe, H., Bekaert, P., 2009,  MNRAS, 397, 341.

Mao, S., Schneider, P., 1998,  MNRAS, 295, 587.

Mao, S., Jing, Y., Ostriker, J.P., Weller, J., 2004,  ApJ, 604L, 5.

Metcalf, R.B., 2002, ApJ, 567, L5

Moore, B., Ghigna, S., Governato, F., Lake, G., Quinn, T., Stadel, J., Tozzi, P. 1999,  ApJ, 524L, 19.

Myers, S.T., Jackson, N.J., Browne, I.W.A., de Bruyn, A.G., Pearson, T.J., Readhead, A.C.S., Wilkinson, P.N., Biggs, A.D., Blandford, R.D., Fassnacht, C.D.,  2003,  MNRAS, 341, 1.

Nair, S., 1998,  MNRAS, 301, 315.

Oguri, M., et al., 2004,  ApJ, 605, 78.

Oguri, M., 2010, PASJ, 62, 1017

Ota, N., et al., 2006, ApJ, 647, 215

Phillips, P.M., et al., 2000, ApJ, 319, L7

Refsdal, S., 1964, MNRAS, 128, 307

Saha, P., Williams, L.L.R., Ferreras, I. 2007,  ApJ, 663, 29.

Schechter, P., Wambsganss, J., 2002., ApJ, 580, 685

Schinnerer, E., 2007, ApJS, 172, 46

Schneider, D.P., 2007, AJ, 134, 102

Sharon, K., et al., 2005,  ApJ, 629L, 73.

Sykes, C.M., Browne, I.W.A., Jackson, N.J., Marlow, D.R., Nair, S., Wilkinson, P.N., Blandford, R.D., Cohen, J., Fassnacht, C.D., Hogg, D.,  1998,  MNRAS, 301, 310.

White, R.L., Helfand, D.J., Becker, R.H., Glikman, Eilat, de Vries, W., 2007,  ApJ, 654, 99.

Williams, L.L.R., Saha, P., 2004,  AJ, 128, 2631.

Xu, D.D. et al., 2009, MNRAS, 398, 1235


\end{document}